\documentclass[conference]{IEEEtran}
\IEEEoverridecommandlockouts

\usepackage{booktabs} 
\usepackage{amssymb}
\usepackage[ruled,commentsnumbered]{algorithm2e}
\usepackage{amsmath}
\usepackage[FIGTOPCAP]{subfigure}
\usepackage{figure}
\usepackage{multirow}
\usepackage{diagbox}
\usepackage{booktabs}
\usepackage{graphicx}
\usepackage{color}
\usepackage{url}

\makeatletter
\renewcommand{\@algocf@capt@plain}{above}
\makeatother

\renewcommand\figurename{Figure}
\newtheorem{definition}{Definition}
\newcommand\tabcaption{\def\@captype{table}\caption}

\begin{document}

\title{Deep Co-investment Network Learning for Financial Assets} 

\author{\IEEEauthorblockN{Yue Wang\IEEEauthorrefmark{1},
Chenwei Zhang\IEEEauthorrefmark{2}, Shen Wang\IEEEauthorrefmark{2}, Philip S Yu\IEEEauthorrefmark{2}, Lu Bai\IEEEauthorrefmark{1}, and Lixin Cui\IEEEauthorrefmark{1}
\IEEEauthorblockA{\IEEEauthorrefmark{1}Central University of Finance and Economics, Beijing, China\\
Email: wangyuecs, bailucs, cuilixin@cufe.edu.cn}\thanks{Corresponding author: Lu Bai (email: bailucs@cufe.edu.cn).}
\IEEEauthorblockA{\IEEEauthorrefmark{2}Department of Computer Science,
University of Illinois at Chicago, Chicago, USA\\
Email: czhang99, swang224, psyu@uic.edu}}
}

\maketitle

\begin{abstract}
Most recent works model the market structure of the stock market as a correlation network of the stocks. They apply pre-defined patterns to extract correlation information from the time series of stocks. Without considering the influences of the evolving market structure to the market index trends, these methods hardly obtain the market structure models which are compatible with the market principles. Advancements in deep learning have shown their incredible modeling capacity on various finance-related tasks. However, the learned inner parameters, which capture the essence of the finance time series, are not further exploited about their representation in the financial fields. In this work, we model the financial market structure as a deep co-investment network and propose a Deep Co-investment Network Learning (DeepCNL) method. DeepCNL automatically learns deep co-investment patterns between any pairwise stocks, where the rise-fall trends of the market index are used for distance supervision. The learned inner parameters of the trained DeepCNL, which encodes the temporal dynamics of deep co-investment patterns, are used to build the co-investment network between the stocks as the investment structure of the corresponding market. We verify the effectiveness of DeepCNL on the real-world stock data and compare it with the existing methods on several financial tasks. The experimental results show that DeepCNL not only has the ability to better reflect the stock market structure that is consistent with widely-acknowledged financial principles but also is more capable to approximate the investment activities which lead to the stock performance reported in the real news or research reports than other alternatives.
\end{abstract}


\section{Introduction}
The market structure \cite{JOFI:JOFI339} is a core issue in economics since it can help the economist understand the dynamic patterns in the market. Generally, the markets can be observed by a group of time series, e.g. the stock market can be treated as a group of varying price and volume time series. Most recent works model the market structure as a correlation network with the statistical methods \cite{DBLP:conf/kdd/JohnsonB15} or methods based on similar pattern matching \cite{DBLP:conf/icdm/SilvaBK16}. However, since many existing works try to model the relations between financial assets with pre-defined patterns without considering the relationship of the stock-level and the market-level financial time series (e.g. the impact of the stock index on the market index trend), it is still challenging to fully address the following aspects:
\begin{itemize}
  \item \textbf{Prior pattern dependency: }Many statistical methods model the market structure as the time series correlation network based on the prior patterns (e.g. Pearson correlation coefficient \cite{DBLP:conf/kdd/JohnsonB15}) designed by experts, and they ignore all the other patterns that are potentially useful. Therefore, their results may bias from the practical rules.
  \item \textbf{Stop word bias \cite{DBLP:conf/kdd/DauK17}: }The truly informative patterns may be overwhelmed by the numerous frequent results which are caused by the regular activities similar to the stop word bias. Therefore, it is challenging to distinguish the regular and the truly interesting patterns for these methods.
  \item \textbf{Evolving patterns: }The market time series system is evolving with the new data, and the co-varying patterns between the time series are also changing. Consequently, it may cause unexpected results when people analyze the time series with the outmoded patterns.
\end{itemize}

To derive a financial market structure that reveals the investment activities between pairwise stocks in the market which is hard to obtain directly, we propose a Deep Co-investment Network Learning method, namely DeepCNL, to learn a stock market structure that reflects co-investment activities (deep co-investment network) via automatically learned non-linear co-investment patterns (deep co-investment patterns). Our study starts from a financial mechanism about the buy-sell imbalance and the assets price. As introduced in the literature \cite{JOFI:JOFI2626}, there is a positive relationship between the buy-sell pressure and the asset price. Therefore, if the transaction prices and volumes of two stocks have similar rise and fall patterns, there is a high probability that these two stocks are invested together by a same group of people. We consider this as the co-investment activity. Since the co-investment activities cannot be observed directly, DeepCNL captures the evidence for them through convolution operators on the stock time series, where the convolution kernels represent the deep co-investment patterns, at a Convolutional Neural Network (CNN) layer. Considering the convolution kernels are learned automatically during the training process, the obtained deep co-investment patterns do not depend on any prior patterns and they can evolve with the new data. With the evidence of the co-investment activities, we represent the stock market as the deep co-investment network that addresses all the co-investment relationships for the stocks in a stock market. DeepCNL is supervised by the rise-fall trends of the index of the whole market and learns the deep co-investment network by the inner parameters of a Recurrent Neural Network (RNN) layer that captures temporal impacts of any pairwise stocks on the market index trends. This step also alleviates the stop word bias issue by supervising the learned results (both the deep co-investment patterns and network) with the market index trends. We verify the effectiveness of the deep co-investment network in addressing the investment activities on the real-world data. Several financial tasks are introduced to interpret the compatibility of the learned deep co-investment networks from DeepCNL with widely acknowledged financial principles.

In summary, the main contribution of this work including:
\begin{itemize}
  \item We propose a Deep Co-investment Network Learning method, namely DeepCNL, to automatically learn a co-investment network, which reflects the intensities of the co-investment activities for any pairwise stocks.
  \item DeepCNL connects the learning of the co-investment network to the rise-fall trends of the market index. The modeling of deep co-investment patterns and the resulting deep co-investment network do not rely on any pre-defined patterns.
  \item We build the deep co-investment network with the learned inner parameters of DeepCNL which reveal the impacts of the stock index on the market index trends. To the best of our knowledge, this paper is the first to study the relation between the co-investment activities and the inner parameters of the deep learning model.
  \item We observe that DeepCNL learns a deep co-investment network that performs consistently with known financial principles about the investment activities.
\end{itemize}

\section{Preliminary}


The assets usually have several attributes, e.g. the observation about a stock contains both its prices and transaction volumes at a given time. Therefore, this work treats all the original time series for stocks as the vector time series.
\begin{definition}
(\textbf{Vector Time Series}) A vector time series $S$ is denoted as a sequence of real-valued vectors, $S=[S(0),S(1),...,S(t),...S(N)]$ is a $\mathbb{R}^{M\times N}$ matrix, where $S(t)=(f_{t,0},f_{t,1},...f_{t,M-1})^T$ is a $\mathbb{R}^M$ vector which represents the values of the $M$ features at the $t-$th time stamp, and $N$ is the length of $S$.
\end{definition}
We simplify the term ``vector time series'' as ``time series'', $S(t)$ as a scalar and $S=(S(0),S(1),...,S(t),...S(N))^T$ as a $\mathbb{R}^N$ vector for brevity in the following part. We follow the method from Lee et al. \cite{DBLP:conf/sdm/LeeKBM17} to concatenate two aligned time series $S_i$ and $S_j$ with length $N$ into a $\mathbb{R}^{2\times N}$ matrix $A_{i,j}$.
\begin{equation}
A_{i,j}=[S_i,S_j]^T,
\end{equation}
where $A_{i,j}$ can also be written as $A_{i,j}=[A_{i,j}(0),A_{i,j}(1),...,$ $A_{i,j}(N)]$, each $A_{i,j}(t)=\langle S_i(t),S_j(t)\rangle$ ($t\in[0,N-1]$) is a dyadic tuple which represents the corresponding values of the time series $S_i$ and $S_j$ at the $t$-th time stamp. We call the matrix $A_{i,j}$ as the \textbf{observation matrix} for $S_i$ and $S_j$.

\begin{definition}
\textbf{(Deep Co-investment Pattern)} Given two time series $S_i$ and $S_j$, let $A_{i,j}$ be their observation matrix, then the deep co-investment pattern between $S_i$ and $S_j$ can be denoted as $\mathbb{R}^{K\times L\times 2}$ tensor $C=\{C^0,C^1,...,C^{K-1}\}$, where each $C^k$ ($k\in[0,K-1]$) is a $\mathbb{R}^{L\times 2}$ matrix which represents the $k$-th co-varying latent pattern of $S_i$ and $S_j$ in the window of size $L$. $K$ is a parameter to decide the number of the different co-varying latent patterns.
\end{definition}

The co-varying latent pattern is a framework to capture the co-varying rules between two time series, and it is a generalized form of many existing correlation measurement methods. In our deep learning model, the co-varying latent patterns are automatically learned during training. The bigger the $K$ is, the more co-varying latent patterns will be collected by our model. According to the ``buy-sell pressure'' principle in the introduction, if two stocks are invested by the same group of investors, their feature time series (e.g. prices, volumes, etc.) will vary analogously (with the similar rise or fall trends). Therefore, the deep co-investment patterns can be used as the filters to capture the clues for two stocks being frequently invested together. We propose the deep co-investment evidence to record the filtered clues.
\begin{definition}
\textbf{(Deep Co-investment Evidence)} Given the observation matrix $A_{i,j}=[S_i,S_j]^T$, and a deep co-investment pattern $C$ with the window of size $L$ for $S_i$ and $S_j$; the deep co-investment evidence $x_{i,j}=[x_{i,j}(0),x_{i,j}(1),...,$ $x_{i,j}(N-L)]$ is a $\mathbb{R}^{K\times (N-L+1)}$ matrix resulted from the convolution between $A_{i,j}$s and all the $\mathbb{R}^{L\times 2}$ co-varying pattern $C^k$ ($k\in[0,K-1]$). Denote vector $x_{i,j}(t)=(x^0_{i,j}(t),x^1_{i,j}(t),...,$ $x^{K-1}_{i,j}(t))^T$, where $x^k_{i,j}(t)$ is a scalar ($k\in[0,K-1]$, $t\in[0,N-L]$) which represents the co-investment degree between $S_i$ and $S_j$ at the $t$-th time stamp of the $k$-th co-varying latent pattern. $x^k_{i,j}(t)$ can be computed as:
\begin{equation}
x^k_{i,j}(t)=C^k*[A_{i,j}(t),A_{i,j}(t+1)...,A_{i,j}(t+L-1)] + B_c,
\end{equation}
where $*$ is the convolution operator and $B_c$ is the bias.
\end{definition}

The deep co-investment evidence records the co-investor activities which follow the deep co-investment patterns. For example, given time series $S_0=(1,1,1,1,1,1)^T$ and $S_1 =(1,1,0,1,1,0)^T$, the observation matrix $A_{0,1}$ is:
\begin{equation}
A_{0,1}=\left[\begin{matrix}
   1 & 1 & 1 & 1 & 1 & 1\\
   1 & 1 & 0 & 1 & 1 & 0
  \end{matrix}\right].
\end{equation}
Let the deep co-investment pattern $C$ be a $\mathbb{R}^{2\times 3 \times 2}$ tensor $C=\{C^0,C^1\}$, where the co-varying patterns $C^0=[(0,0)^T,(0,0)^T,(0,1)^T]$ and $C^1=[(1,1)^T,(1,1)^T,(1,0)^T]$ are the $\mathbb{R}^{3 \times 2}$ matrices, and suppose the bias is zero. Then the deep co-investment evidence $x_{0,1}$ is:
\begin{equation}
x_{0,1}=\left[\begin{matrix}
   0 & 1 & 1 & 0 \\
   5 & 4 & 4 & 5
  \end{matrix}\right].
\end{equation}

In this example, $x_{0,1}$ collects the evidence both for the co-varying latent pattern $C^0$ and $C^1$ in $C$.

We define the deep co-investment network to describe the relationships for all dyadic combinations of the stocks being invested together in a market.
\begin{definition}
\textbf{(Deep Co-investment Network)} The deep co-investment network is a weighted undirected graph $G=\langle V,E,W\rangle$, where the node set $V$ represents all the stocks in the market, and the edge set $E$ represents the deep co-investment relationships between any two stocks. The weight vector $W=(w_{0,1},...,w_{i,j},...,w_{|V|,|V|-1})^T$ represents the co-investment intensities between the two stocks, where a large weight, $w_{i,j}$ $(\forall i,j\in V, i \neq j)$ indicates the trends that the investors spend more money on the stock $i$ and $j$ together.
\end{definition}

Generally, the transactions between the stocks and investors can be modeled as a bipartite graph. In addition, if the data are available, an oracle co-investment network can be computed by a one-mode projection with the transactions between the stocks and investors. However, since the transaction records are hard to collect for many reasons (e.g. legal, privacy, or technical issues), we construct the deep co-investment network indirectly by the deep co-investment evidence and the market index data.

The rise or fall of the market index is caused by the aggregated investment activities from the stocks in that market. Consequently, given the deep co-investment evidence for all the dyadic stock combinations, the deep co-investment network learning problem can be formalized as follows.

\begin{definition}
\textbf{(Deep Co-investment Network Learning)} Suppose $V$ is the set of all the stocks in a market, all the stock time series in $V$ are aligned with length $N$, and the window size for the deep co-investment pattern is $L$. $X=\{x_{0,1},x_{0,2}, ..., x_{|V|-1,|V|}\}$ is a $\mathbb{R}^{\binom{|V|}{2}\times K\times(N-L) }$ tensor, where $x_{i,j}$ is the deep co-investment evidence for the stock time series $S_i$ and $S_j$ in $V$. $X(t) = [x_{0,1}(t),x_{0,2}(t),...x_{|V|-1,|V|}(t)]$ is a $\mathbb{R}^{\binom{|V|}{2}\times K}$ matrix, and it is the deep co-investment evidence for all the dyadic combinations of the stocks at the $t$-th time stamp. Let $Y$ be a binary time series which representing the rise or fall trends of the market index, our goal is to estimate the optimal weight vector $W$ for the deep co-investment network $G=\langle V,E,W\rangle$ which can maximize the probability of computing $Y$ given the deep co-investment evidence. It can be formalized as the following.
\begin{equation}
\arg\max_{W}{\prod_{t=0}^{N-L}{p(Y(t+L)|R(t),...,R(t+L-1))}},
\end{equation}
where $Y(t)$ is the value of $Y$ at the \emph{t}-th time stamp. $R(t)$ is a $\mathbb{R}^{\binom{|V|}{2}}$ vector and it can be computed as the follow equation:
\begin{equation}
R(t)=WX(t)^T+B_R,
\end{equation}
where $B_R$ is a $\mathbb{R}^{\binom{|V|}{2}}$ bias vector. 
\end{definition}

With the aforementioned notations, the co-investment evidence $x_{i,j}(t)$ records the co-investment activities by the investors and $w_{i,j}$ is the co-investment intensities on the stock $i$ and $j$ by the investors. While the weight $w_{i,j}$ is directly related to the money spent on the stock $i$ and $j$ by the investors with the fixed co-investment evidence. Furthermore, as the market index (aggregate index price movement) is directly correlated to the total fund flow (or invested money) \cite{WARTHER1995209}, it is correlated to the sum of co-investment intensities (or weights) for all the stocks. Therefore, only the correct co-investment intensities for all the corresponding pairs of the stocks lead to the real market index when the corresponding co-investment evidence is converged after the optimization process, and thus, supervised by the market index, we can learn the effective co-investment intensity for each pair of stocks.

Since solving this problem would require trying different weight values on all dyadic combinations of the time series, it is an NP-hard combinatorial optimization problem. We propose a deep learning model solution by applying the GPU to speed up the computation.

\section{Our Framework}

We propose the Deep Co-investment Network Learning (DeepCNL) framework,
\begin{figure}[h]\vspace{-0.1in}
\centering
\caption{The framework of DeepCNL}
\includegraphics[width=3.35in]{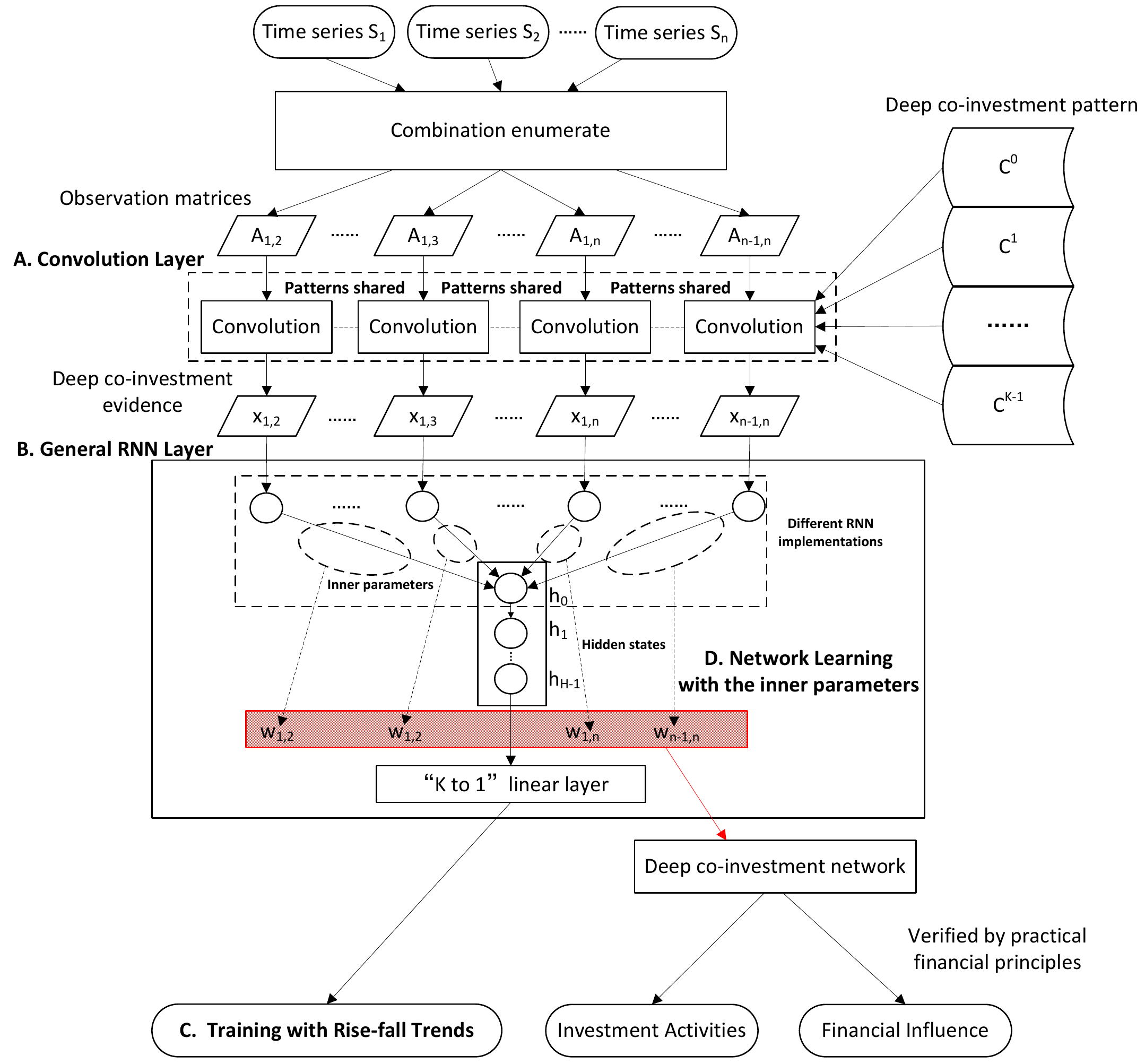}
\end{figure}
 as shown in Figure 1, the DeepCNL model combines a convolution layer, a general RNN layer, and a rise-fall trends supervision layer. (1) The convolution layer enumerates all the dyadic combinations of the time series and concatenate every pair of time series $i$ and $j$ ($\forall i,j\in V, i\neq j$) to a set of the observation matrices $D=\{A_{0,1},A_{0,2},...,A_{|V-1|,|V|}\}$, and computes all the deep co-investment evidence for the observation matrices through a convolution layer; (2) the general RNN layer gets all the deep co-investment evidence input into a $\binom{|V|}{2}$-channel RNN with each deep co-investment evidence $x_{i,j}$ corresponding a specific input channel, and outputs a single predicted real-valued sequence; (3) In the rise-fall trends supervision layer, DeepCNL measure the loss between the predicted sequence from the RNN layer and the target rise-fall sequence $Y$ and apply the back propagation methods to optimize the parameters. (4) DeepCNL learns the deep co-investment network with the learned inner weights of RNN after the training with the supervision of the rise-fall trends. The obtained deep co-investment network can be applied to various financial tasks.

\subsection{Convolution Layer}

The convolution layer \cite{deeplearning2015} is a critical part of DeepCNL to trace the deep co-investment patterns between two time series. Since our model aims to capture a full spectrum of deep co-investment patterns and computes the result sequences which should align to the market index time series with the rise-fall pattern, it uses the convolution layer directly without any down sampling methods \cite{6738831}.

\begin{algorithm}[h]\scriptsize
\DontPrintSemicolon \KwData{stock set of the market $V$, deep co-investment pattern $C$, bias $B_c$}
\KwResult{deep co-investment evidence tensor $X=\{x_{0,1},x_{0,2}, ..., x_{|V|-1,|V|}\}$}
\Begin {
$X\leftarrow \phi$\\
$D=\{A_{0,1},A_{0,2},...,A_{|V-1|,|V|}\}$ $\leftarrow$enumerate($V$)\\
\For{$\forall A_{i,j}\in D$}{
\For(\tcc*[f]{analyze all $K$ co-varying latent patterns}){$\forall C^k\in C$}{
\For{$t=1$ \KwTo $N$}{
\tcc{Apply equation (2) to compute $x^k_{i,j}$}
$x^k_{i,j}(t)\leftarrow C^k*[A_{i,j}(t),...,A_{i,j}(t+L-1)] + B_c$\\
}
}
$X\leftarrow$concatenate($X,A_{i,j}$)
}
\textbf{Output} $X=\{x_{0,1},x_{0,2}, ..., x_{|V|,|V|-1}\}$
}
\caption{Convolution computation process}
\end{algorithm}

With the notations in Definition 2, the computation process of our convolution layer is designed in Algorithm 1, where the function ``enumerate($V$)'' generates all the dyadic combinations from $V$, and ``concatenate$(*)$'' appends the newly computed $A_{i,j}$ to the end of the first dimension of $X$ at each iteration. One should note that all the convolutions share the same $C^k$ during different sliding time windows. Therefore, this methods captures the global co-investment structure for all the stocks with the same deep co-investment pattern. Since the algorithm 1 needs to enumerate the complete dyadic combinations, its complexity is $O(|V|^2\times|C|\times N)$.

\subsection{General RNN Layer}
The deep co-investment network learning is a ``n to 1'' sequence learning task which means to learn a single-feature sequence given a multi-feature sequence (such as the vector time series). This conforms to the scene which predicts the rise or fall trends of the market index based on the former observations of the stock indices in a market. As illustrated in the Figure 1, we apply a general multi-feature recurrent neural network framework for the sequence learning task. The RNN framework has the flexibility to adopt the basic RNN \cite{LUKOSEVICIUS2009127}, the long-short memory recurrent neural network (LSTM) \cite{iet:/content/conferences/10.1049/cp_19991218}, the gated recurrent neural network (GRU-RNN) \cite{DBLP:journals/corr/ChungGCB14} or any other RNNs according to the tradeoff of performance and accuracy. Our RNN layer consists of the RNN framework with $\binom{|V|}{2}$ input channels, and it computes a score that indicates the rise-fall trend with the process in Algorithm 2,
\begin{algorithm}[h]\scriptsize
\DontPrintSemicolon \KwData{deep co-investment evidence $X=\{x_{0,1},x_{0,2}, ..., x_{|V|-1,|V|}\}$, initial parameters for RNNs, weight $W$ for the deep co-investment network and bias $B_R$, window size $L$, and the weights $\alpha$ and bias $\beta$ for the ``$K$ to 1'' linear layer}
\KwResult{A rise-false score time series $Y'$}
\Begin {
$Y'\leftarrow \phi, h(L-2)\leftarrow 0$ \tcc*[r]{initialize the output and hidden state}
\For{$\forall t\in [L-1, N-1]$}{
Out$(t),h(t)\leftarrow$ RNN$(X(t),h(t-1))$\\

$Y'(t)\leftarrow \alpha$Out$(t)^T+\beta$ \tcc*[r]{the "$K$ to 1" linear layer}
$Y'\leftarrow concatenate(Y',Y'(t))$\\
}
\textbf{Output} $Y'=(Y'(L-1),Y'(L),...,Y'(N-1))^T$
}
\caption{RNN computation process}
\end{algorithm}
where the function concatenate$(*)$ is the same function as Algorithm 1 to concatenate the vectors, and the function RNN$(X(t),h(t-1))$ is a general RNN framework which computes the output value from the new input $X(t)$ and its last hidden state $h(t-1)$. Since the deep co-investment evidence tensor contains the evidence for $K$ co-varying latent patterns, the RNN output is a $\mathbb{R}^{K}$ vector. Therefore, we use a ``$K$ to 1'' linear layer to convert the RNN output to a scalar output $Y'(t)$, and $Y'(t)$ is the score which represents the rise-fall trend considering all the deep co-investment evidence in the former window. The RNN in Algorithm 2 needs to process the deep co-investment evidence for all the time stamps, and thus its time complexity is $O(N)$.

\subsection{Training with Rise-fall Trends}
We connect the deep co-investment network learning with the rise-fall trends of the market index with the softmax \cite{Bridle1990} method. We preprocess the target market index into a binary time series $Y=(Y(0),Y(1),...,Y(N))^T$, where $Y(t)=0$ or 1, 1 and 0 represents the rise and fall trend of the market index from the last value respectively, and the classification score $g_t$ of $Y'$ can be computed by:
\begin{equation}
g_t=\langle \frac{e^{Y'(t)}}{1+e^{Y'(t)}},\frac{1}{1+e^{Y'(t)}}\rangle,
\end{equation}
where $t$ is the time stamp, the element $g_t(0)=\frac{e^{Y'(t)}}{1+e^{Y'(t)}}$ and $g_t(1)=\frac{1}{1+e^{Y'(t)}}$ represents the degree of rise and fall trends respectively. Our loss function is defined as follows.
\begin{equation}
Loss(Y',Y)=\frac{1}{N}\sum_{t=0}^{N}\log(\frac{e^{g_t(Y(t))}}{e^{g_t(0)}+e^{g_t(1)}})+\lambda||\theta||_F,
\end{equation}
where $\lambda||\theta||_F$ is the Frobenius norm for all the parameters $\theta$. It is convenient to compute the derivation to this loss, and thus, we apply the back propagation methods with Adam \cite{DBLP:journals/corr/KingmaB14} optimizer to train DeepCNL.

From Equation (5) and (6), one can note that $Y'$ can also be computed through a map of $WX(t)^T+B_R$, where the $X(t)$ is the deep co-investment evidence at the $t$-th time stamp ($t\in[0,N-L]$), $W$ is the weights for the deep co-investment network, and $L$ is the size for the time window. Since $Y'$ can also be obtained by Algorithm 2 of DeepCNL, we can infer that our loss function actually helps optimizing the weight $W$ during the training of DeepCNL. The experiments shows that this inference is correct, and the weight $W$ estimated with the learned inner parameters of DeepCNL positively related to the real co-investment relations for the stocks.

\subsection{Network Learning}

We use LSTM \cite{NIPS2014_5346} to estimate the weight $w_{i,j}$ ($\forall i,j\in V$) for the deep co-investment network in this work. This process can also use any different version of RNN according to the application.
The output $h(t)$ of LSTM is computed as:
\begin{equation}
h(t)=o(t)\tanh(c(t)),
\end{equation}
where the $o(t)$ is the output gate and $c(t)$ is the cell value, and they can be calculated as:
\begin{equation}
o(t)=\delta(W^{io}x(t)+B^{io}+W^{ho}h(t-1)+B^{ho}),
\end{equation}
\begin{equation}
c(t)=f(t)c(t-1)+i(t)g(t),
\end{equation}
where function $\delta(*)$ is the $sigmoid$ function, $W^{io}$ is the weights from the input to the output gate, and $W^{ho}$ is the weights between the hidden states. $B^{io}$ and $B^{ho}$ are the corresponding biases. The $f(t)$ and $i(t)$ are the forget gate and the input gate respectively, and $g(t)$ is the hidden representation for the inputs. They can be obtained by the following equations:
\begin{equation}
f(t)=\delta(W^{if}x(t)+B^{if}+W^{hf}h(t-1)+B^{hf}),
\end{equation}
\begin{equation}
i(t)=\delta(W^{ii}x(t)+B^{ii}+W^{hi}h(t-1)+B^{hi}),
\end{equation}
\begin{equation}
g(t)=\tanh(W^{ig}x(t)+B^{ig}+W^{hg}h(t-1)+B^{hg}),
\end{equation}
where $W^{if}$, $W^{ii}$ and $W^{ig}$ are the weights from the input to the forget gate, input gate and the input hidden representation respectively. $B^{if}$, $B^{ii}$ and $B^{ig}$ are the corresponding biases. $W^{hf}$, $W^{hi}$ and $W^{hg}$ are the weights between the hidden states. $B^{hf}$, $B^{hi}$ and $B^{hg}$ are the biases. We design the LSTM version weight as: 
\begin{equation}
w_{i,j}=\sum_{l=0}^{H-1}(w^{ii}_{i,j}(l)+w^{ig}_{i,j}(l)+w^{io}_{i,j}(l)),
\end{equation}
where $w^{ii}_{i,j}(l)$, $w^{ig}_{i,j}(l)$, and $w^{io}_{i,j}(l)$ are the weights of the input gate, the input hidden representation, and the output gate for $x_{i,j}$ at the $l$-th hidden state on the LSTM.

\textbf{Discussion about the weight estimation.} As one can observe from in Equation (15), the weights for the deep co-investment network are computed by adding up all the weights of the new input gates (e.g. the input to hidden, input gate, etc.), since these weights collect the positive impacts of the deep co-investment evidence on the market index. Thus the estimated weights for the deep co-investment network indicate the total positive impacts from the deep co-investment evidence to the market index. Then the final estimated weight $w_{i,j}$ reveals the impact of $x_{i,j}$ (for stock $i$ and $j$) on the market index sequence. Therefore, the estimated weights will keep a relative deterministic partial order after the training by referring to their corresponding impacts. We verify this intuitive inference in the experiment.

\textbf{Network generation.} As it is described in the former sections, we obtain a weight for every dyadic combination of the stocks. If the final network gets too many edges, it will be a useless dense graph. This is a common issue in the graph learning tasks \cite{DBLP:conf/icml/JebaraWC09}. We design a graph generator framework, applying the ``Occam's Razor law'', we first sort all the combinations by the weights in the descending order. And then, we add a small proportion, which we called rare ratio $\gamma$, of edges into the final graph. This is also a similar process as the Benjamini-Hochberg procedure \cite{doi:10.3102/10769986027001077} which aims to alleviate the multiple comparisons problem for a set of statistical tests when treating the weight estimation for each edge as an independent statistical test.


\section{Experiments and Discussion}

\subsection{Datasets}

We compare our methods with other existing methods on the S\&P 500 dataset. Its details are shown in Figure 2.
\begin{figure}[htbp]
\begin{minipage}[t]{0.48\linewidth}
\centering
\tabcaption{Dataset statistics}
\scalebox{0.65}[0.65]{
\begin{tabular}{ll}
\toprule
Property&S\&P500\\\midrule
Total instants&851,264\\
Company number&470\\
Combination number&110,215\\
Feature number&5\\
Time stamps&1,200\\
Start date&2010-01-04\\
End date&2016-12-30\\ \bottomrule
\end{tabular}
}
\end{minipage}
\begin{minipage}[t]{0.5\linewidth}
\centering
\caption{Coverage comparison}
\includegraphics[height=1.05in]{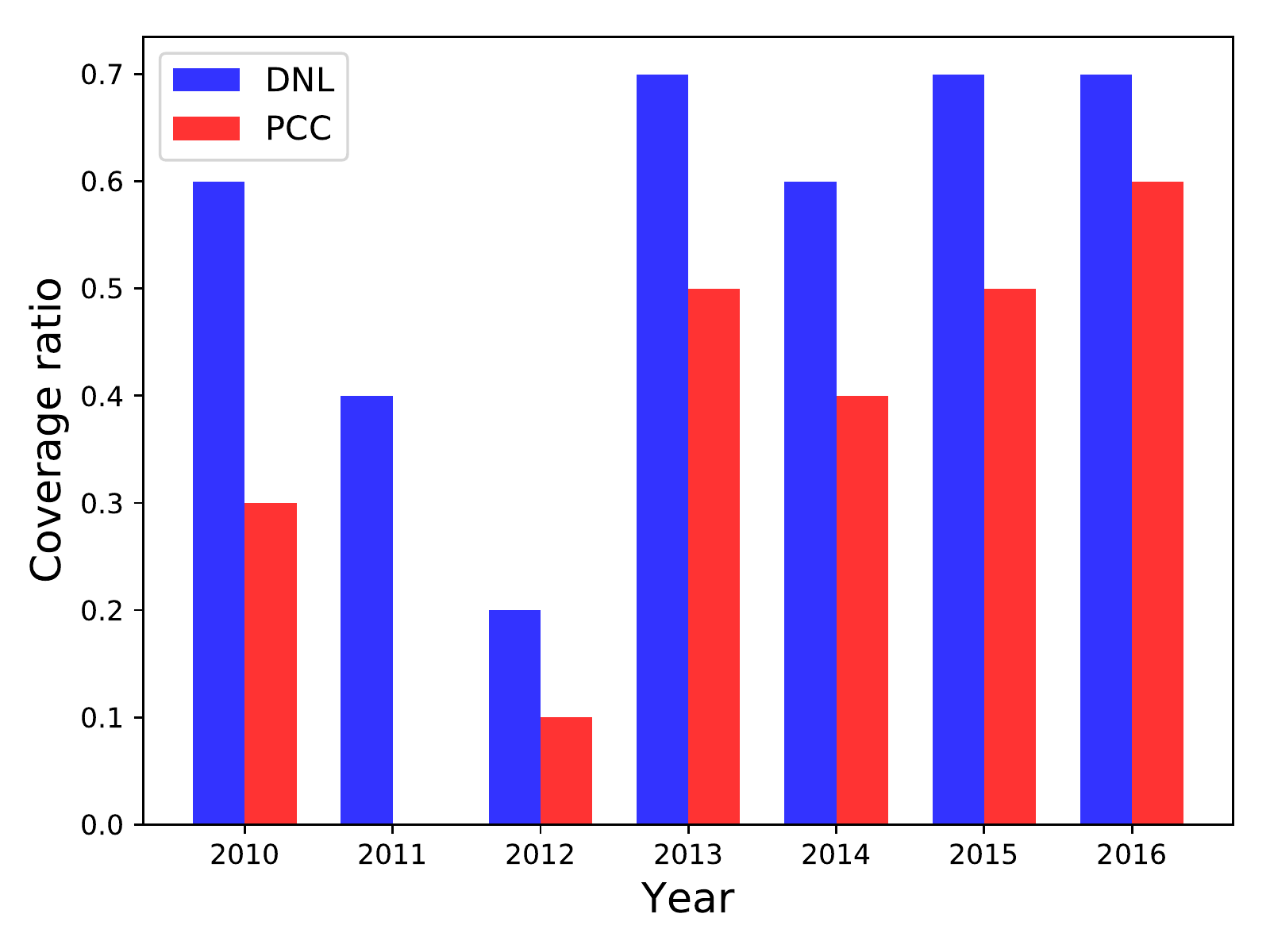}\vspace{-0.1in}

\end{minipage}
\end{figure}\vspace{-0.1in}

This dataset is obtained from a competition targeting on the stock price predicting\footnote{https://www.kaggle.com/dgawlik/nyse}. This dataset of S\&P 500 consists of 470 famous companies since the varying of the components over the years. The 5 features are ``open, high, low, close'' prices and the volumes for the stocks. We use the time series of the ticker ``SPY'' as the target $Y$ for this ETF precisely tracks the market index of S\&P 500. The combination number in Figure 2 is the number of dyadic combinations for the stocks in a market. One can note that, although the original data is not big, the combination number is over 100,000. This means that our model needs to apply the learning process through $110,215$ time series where each has over $1,200$ timestamps (6 years). To capture both long-term and short-term dynamics, we choose one year as the interval and apply the learning process of DeepCNL model on data within one year.

\subsection{Experiment Settings and Benchmark}

Our experiments consist of five parts. First, we verify the effectiveness of DeepCNL on the task about the investment density. Second, we compare DeepCNL with other existing methods on the tasks related to the financial influence. Third, we show the capability of DeepCNL to capture the annual evolving patterns from real data. What's more, we analyze the training process and the correctness of different DeepCNL implementations. Last but not least, we analyze the scalability of all mentioned methods. Since the result in $\uppercase\expandafter{\romannumeral4}.F$ shows that DeepCNL implemented with LSTM performances the best of all the RNNs, we use the LSTM version of DeepCNL for all the comparison experiments.

In our experiments, we set the number of hidden state to 256, the number of hidden layers to 2 for all different RNN implementations of DeepCNL, and we set the number of co-varying latent patterns $K$ to 16 for DeepCNL. The methods used for comparison including:
\begin{itemize}
  \item Pearson Correlation Coefficient (PCC). The PCC is the most common method to analyze the correlation relationships for the time series \cite{DBLP:conf/kdd/JohnsonB15}.
  \item Dynamic Time Warping (DTW) \cite{DBLP:conf/kdd/BerndtC94}. DTW is a dynamic similarity measure method on the time series. It can compare the time series adaptively and does not need to select the features or align the time series manually.
  \item Visibility graph and WL kernel-based method (VWL). We propose VWL to compare the time series through the graph kernel methods. Visibility graph \cite{Lacasa01042008} can transform a single time series as a graph. Based on this idea, we transform all the time series into the graphs, and we apply a graph kernel, Weisfeiler-Lehman kernel \cite{DBLP:conf/nips/ShervashidzeB09}, to compare the similarities between the stocks.
\end{itemize}

We preprocess the original data with a min-max normalization method. The edge weights of the found networks for each method (we simplified DeepCNL as DNL in all the experiments) are computed as the following: DNL uses the deep co-investment network weights; DTW computes the DTW distance $d$ between two time series, and then use the $1/(d+1)$ as the edge weights; PCC obtains the weights by computing the Pearson correlation coefficient between two time series (with p-value$<0.01$); VWL generates the visibility graph for each time series, and computes the similarities between the graphs through WL kernel as the edge weights. Our prototype system\footnote{https://github.com/hkharryking/deepcnl} is implemented with Pytorch.

\subsection{Analysis on Investment Intensity}
Since the edges of the deep co-investment network reveal the relationships that two stocks are invested together by a same group of people, the investment intensity for a group of stocks can be measured by the edge density of the corresponding deep co-investment network. To verify if the learned networks capture the practical investment activities, we compare the edge densities of the subgraphs from the learned networks with DeepCNL and PCC. Each subgraph is related to a ETF which contains a subset of all the component stocks of S\&P 500. The ETFs we used including: the Guggenheim Russell top 50 mega ETF (ticker: ``XLG'') containing the select top 50 big companies in S\&P 500; the S\&P 100 index (ticker: ``OEX'') containing the select top 100 big companies in S\&P 500; ishares Russell top 200 index (ticker: ``IWL'') containing the select top 200 big companies in S\&P 500. The XLG's component set is the subset of OEX's component set, and the OEX's component set is the subset of IWL's component set. That is, if people invest OEX, they will also invest the component stocks in the XLG, and if people invest IWL, they will also invest the component stocks in  XLG and OEX. Since the deep co-investment describes the relation when two stocks are invested by same group of people, the bigger the edge density for one subgraph, the more invested activities are captured in it. We compare the average edge densities for the subgraphs from the annual deep co-investment networks obtained by 5 independent trials for each year's data, and we set the $\gamma =0.001$ for DeepCNL and PCC. The results are shown in Figure 4, where the postfixes ``i'',``g'',``o'', and ``f'' refers to the weights of the input gate, the input hidden representation, the output gate and forget gate of the LSTM respectively. For example, DNL-igo refers to learning the network weight of the deep co-investment network with Equation (15). Figure 4 also illustrates the results of other implementations of DeepCNL with adding up the weights of the corresponding gates. It can be clearly observed that the edge densities for the subgraphs of XLG are the biggest, and the edge densities for the subgraphs of IWL are the smallest for the results of DNL-igo. Since the investment activities on a ETF will accumulate to its subset ETFs, the edge density for the subgraph of ETF will be smaller than the edge density for any subgraph of its subset ETF. \textbf{The edge density result of DNL-igo consistently coincides with the order ``XLG$>$OEX$>$IWL''}, which justifies the effectiveness of using the inner parameters of RNN to represent the co-investment relations in Section $\uppercase\expandafter{\romannumeral3.D}$. Therefore, DNL-igo learns the deep co-investment network which reveals the real market structure and we use it in all the remaining experiments (we omit the suffix ``igo'').
\begin{figure*}[ht]
  \centering
    \caption{Comparison on the edge densities. The solid lines are the results which conform to the order ``XLG$>$OEX$>$IWL'' and the dash lines are the results which violate the order ``XLG$>$OEX$>$IWL''.}
  \subfigure[DNL-igo]{
    \label{fig:subfig:a} 
    \includegraphics[width=1.35in]{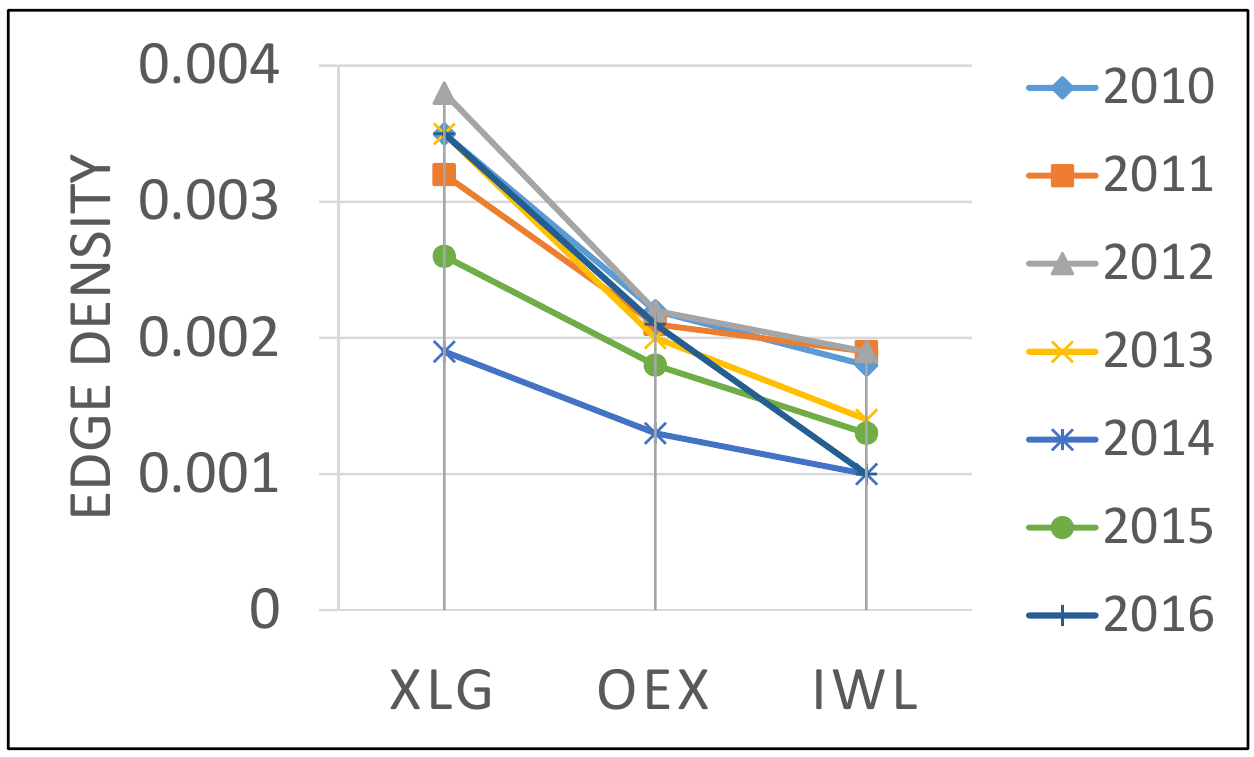}}
  \subfigure[DNL-igof]{
    \label{fig:subfig:b} 
    \includegraphics[width=1.28in]{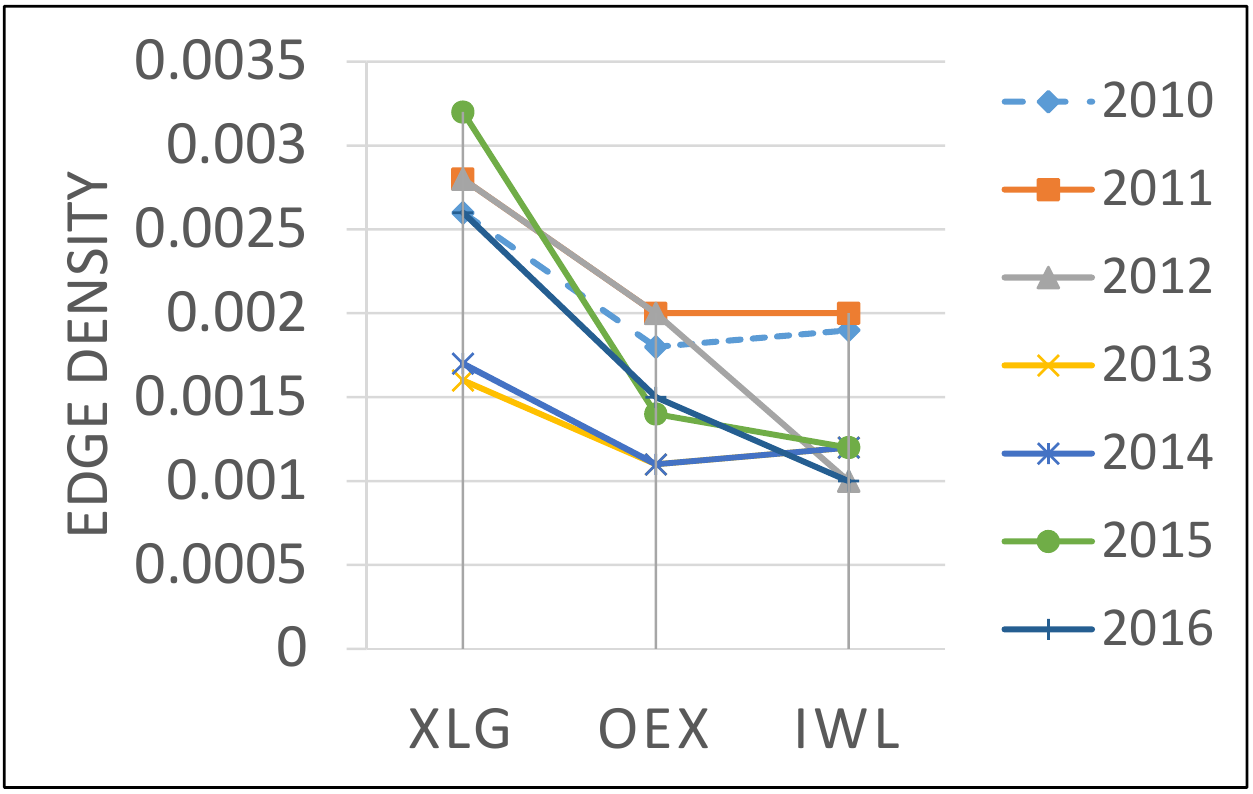}}
     \subfigure[DNL-io]{
    \label{fig:subfig:c} 
    \includegraphics[width=1.31in]{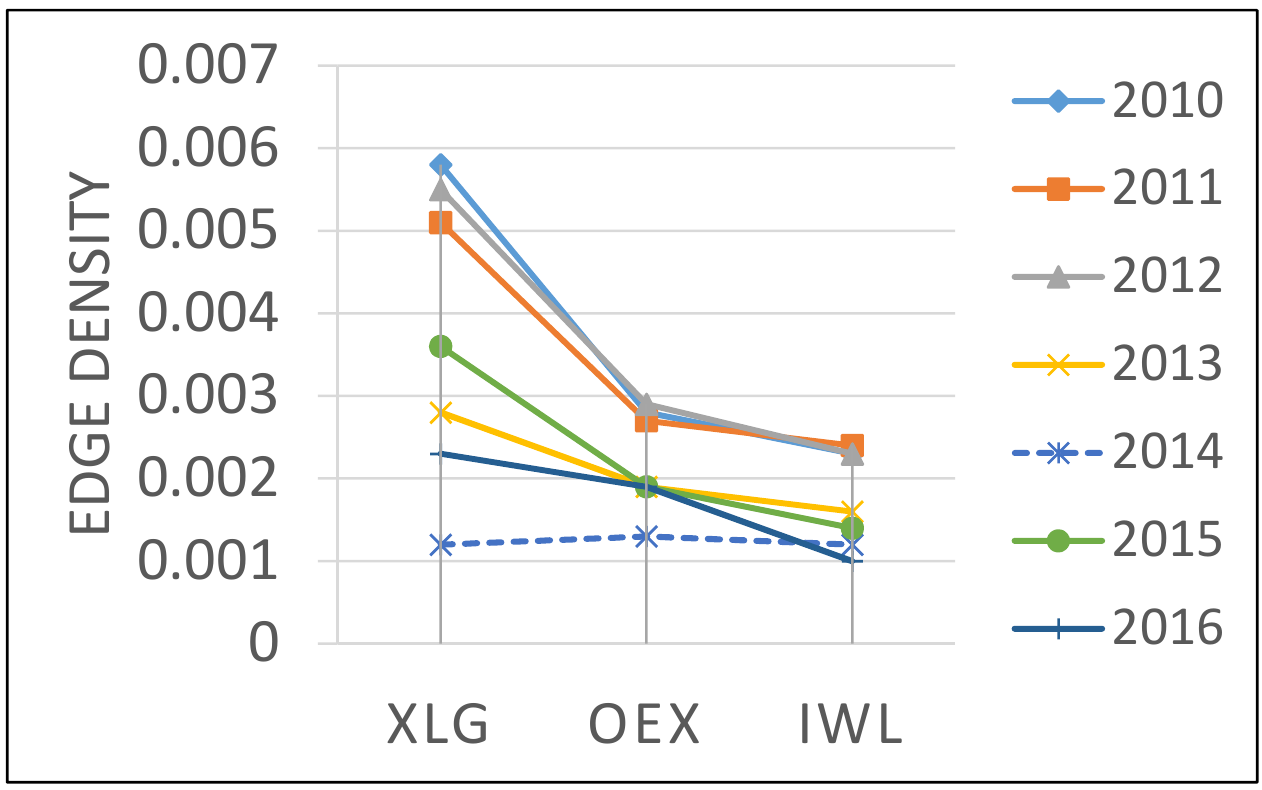}}
     \subfigure[DNL-g]{
    \label{fig:subfig:d} 
    \includegraphics[width=1.28in]{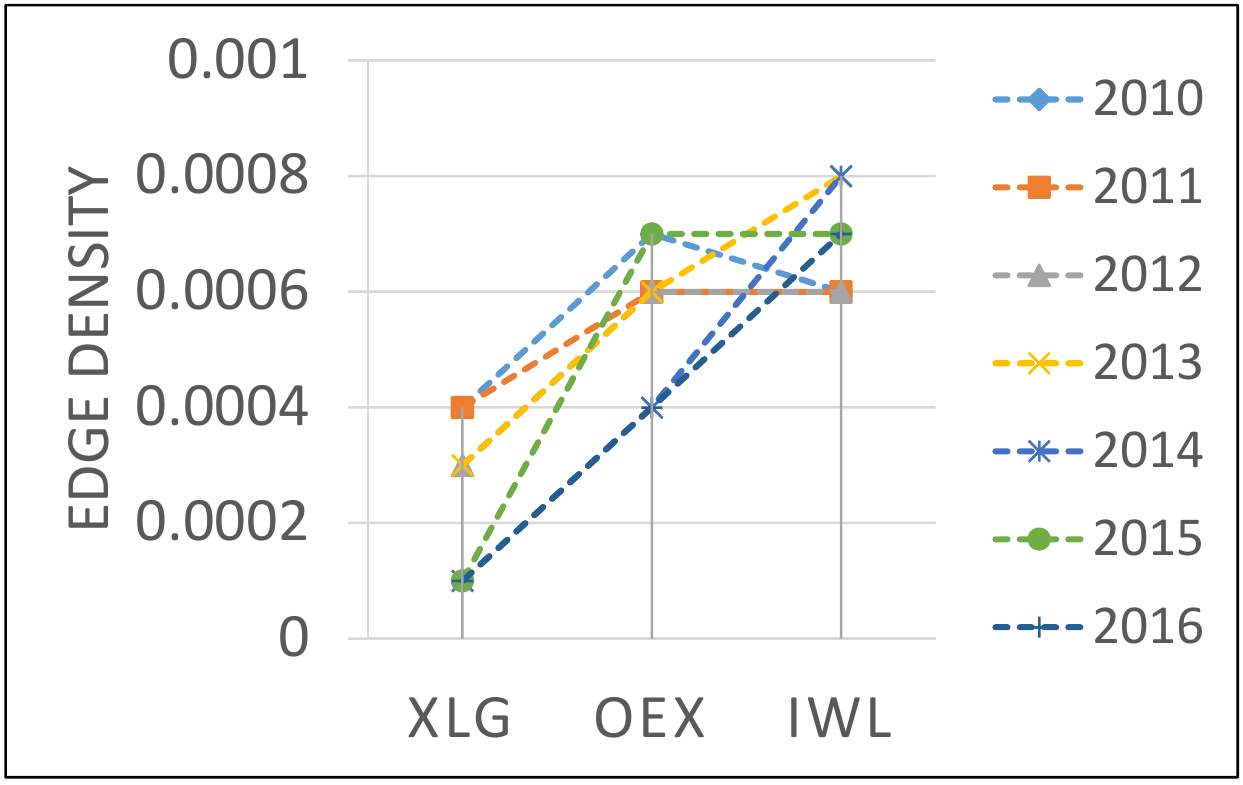}}
    \subfigure[PCC]{
    \label{fig:subfig:d} 
    \includegraphics[width=1.3in]{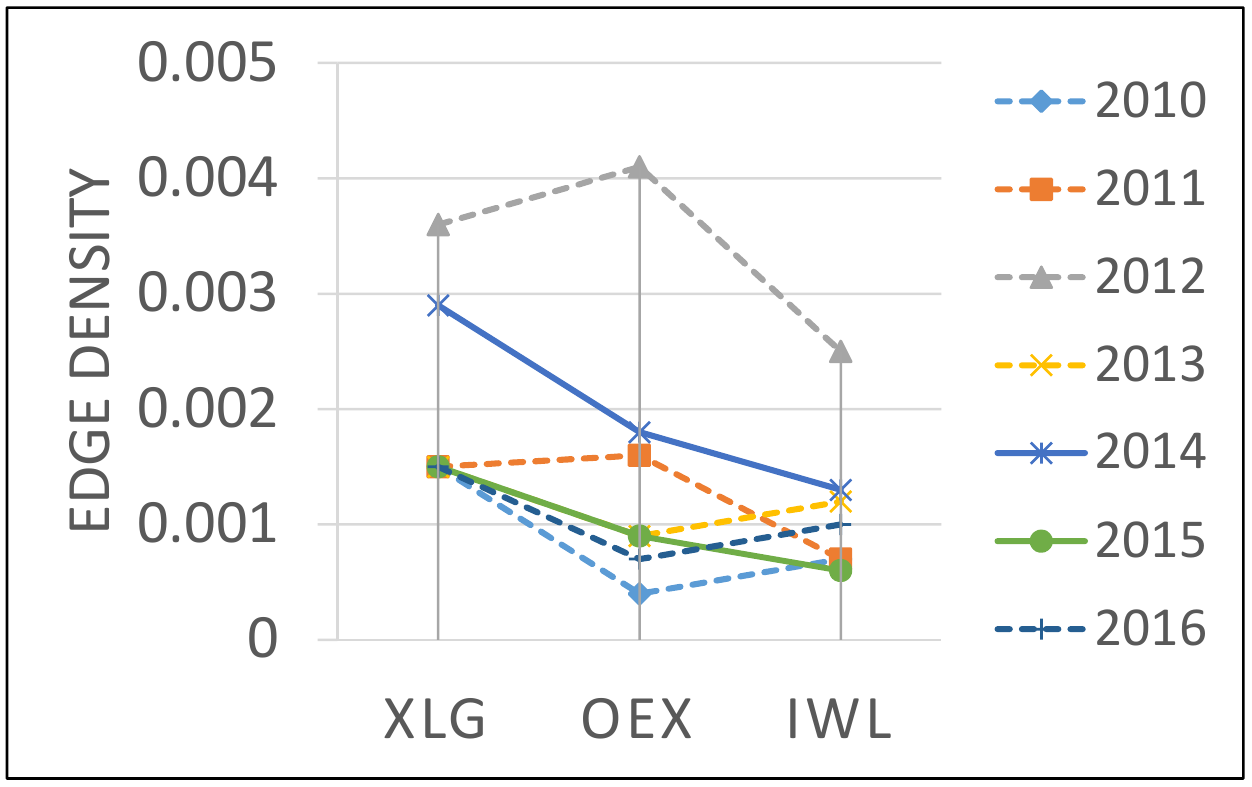}}
  \label{fig:subfig} \vspace{-0.1in}
\end{figure*}

\subsection{Consistency with Practical Financial Principles}

We provide two subtasks to further verify if the learned networks capture the practical financial principles: (1) we analyze the financial influence of the top 10 degree nodes from the learned networks, and (2) whether the learned networks captures the investment activities which lead to the stock performance in the real news.

Since the S\&P 500 index is capitalization-weighted, its component stocks with higher market capitalization have the bigger impact on the index value than those with small market capitalizations\footnote{https://en.wikipedia.org/wiki/S\&P\_500\_Index\#Weighting}. Therefore, we use the average market capitalization as the metric to measure the financial influence of the found stocks. In social network field, the high-degree method is a non-stochastic metric for identifying the influence of a selected node set \cite{Kempe:2003:MSI:956750.956769}. That is, given a fixed node set in a social network, its influence is measured by the sum of the degrees of its nodes. Consequently, our first comparison is made among the average market capitalizations of the 10 top degree nodes of the networks respectively. Since DTW and VWL are not able to scale up to data of this scale, we compare the methods by first sorting all the stock symbols in alphabetic order, and then comparing the four methods with the first 50, 100, 150, and 200 stock time series. We also compare DTW and DNL on all the 470 stocks. We learn the networks with the annual split data during the 6 years and compute the top degree nodes from the learned networks. We set the rare ratio $\gamma=0.02$ for the experiments with 200 or fewer stocks, and $\gamma=0.002$ for the experiments with 470 stocks. The results are reported in Table $\uppercase\expandafter{\romannumeral1}$.
 \begin{table}[h]\scriptsize
\centering
\caption{Comparison on market cap. (USD bn)}\vspace{-0.1in}
\begin{tabular}{ccccc}
\toprule
Stock num.&DNL&PCC&DTW&VWL\\\midrule
50&\textbf{134.0$\pm$39.3}&90.1$\pm$49.4&35.8$\pm$5.3&57.1$\pm$42.1\\
100&\textbf{132.3$\pm$38.6}&70.3$\pm$39.1&42.9$\pm$11.8&50.3$\pm$40.1\\
150&\textbf{169.0$\pm$61.3}&62.1$\pm$32.7&35.4$\pm$7.9&NA\\
200&\textbf{176.1$\pm$56.9}&63.0$\pm$34.4&38.4$\pm$6.9&NA\\
470&\textbf{223.1$\pm$67.7}&67.0$\pm$35.8&NA&NA\\
\bottomrule
\end{tabular}
\end{table}
We can observe from Table $\uppercase\expandafter{\romannumeral1}$ that \textbf{DNL beats all the other methods on the average market capitalizations}, while \textbf{PCC performs the best among the baselines}. Furthermore, the influence of the top stocks from PCC is decreasing with the number of stocks, while the influence of the top stocks from DNL is increasing. This shows that DNL captures the deep co-investment patterns correctly, and it alleviates the stop word bias by getting more global rules than others. To further compare the DNL with the best baseline PCC, we collected the news about ``annual best 10 performing S\&P 500 stocks''\footnote{e.g. http://www.nasdaq.com/article/5-best-performing-sp-500-stocks-of-2014-analyst-blog-cm425786} during the year 2010 to 2016. These news reports ranked the stocks according to their returns. The performance of the stocks has a positive relationship with the real-world investment activities. That is, the higher frequency a stock is invested, the better its performance will be. Therefore, we use the ranked lists of the top-performing stocks as another benchmark to test that how similar the obtained networks conform with the investment activities from people. We list the covered (blue) stocks by the learned network of DNL in Table $\uppercase\expandafter{\romannumeral2}$,
 \begin{table}[h]\scriptsize
\centering
\caption{The covered (blue) annual top 10 performing stocks by the learned network of DeepCNL}\vspace{-0.1in}
\scalebox{0.88}[0.88]{
\begin{tabular}{cccccccc}
\toprule
Rank&2010&2011&2012&2013&2014&2015&2016\\
\midrule
1&\color{blue}{NFLX}&\color{blue}{COG}&HW&\color{blue}{NFLX}&\color{blue}{LUV}&\color{blue}{NFLX}&\color{blue}{NVDA}\\
2&\color{blue}{FFIV}&EP&DDD&\color{blue}{MU}&\color{blue}{EA}&\color{blue}{AMZN}&\color{blue}{OKE}\\
3&\color{blue}{CMI}&\color{blue}{ISRG}&REGN&BBY&EW&\color{blue}{ATVI}&\color{blue}{FCX}\\
4&\color{blue}{AIG}&MA&LL&\color{blue}{DAL}&\color{blue}{AGN}&\color{blue}{NVDA}&CSC\\
5&ZION&BIIB&\color{blue}{PHM}&CELG&MNK&CVC&AMAT\\
6&\color{blue}{HBAN}&HUM&MHO&\color{blue}{BSX}&\color{blue}{AVGO}&\color{blue}{HRL}&\color{blue}{PWR}\\
7&AKAM&\color{blue}{CMG}&AHS&\color{blue}{GILD}&GMCR&\color{blue}{VRSN}&\color{blue}{NEM}\\
8&\color{blue}{PCLN}&PRGO&VAC&YHOO&\color{blue}{DAL}&\color{blue}{RAI}&\color{blue}{SE}\\
9&WFMI&OKS&S&\color{blue}{HPQ}&RCL&SBUX&BBY\\
10&Q&\color{blue}{ROST}&EXH&\color{blue}{LNC}&\color{blue}{MNST}&FSLR&\color{blue}{CMI}\\
\bottomrule
\end{tabular}
}
\end{table}
and we compare the coverage of the best 10 performing stocks for the stocks in the obtained networks of DNL and PCC in Figure 3. We can see from Figure 3, \textbf{DNL covers more top-performing stocks} than PCC, and thus DNL captures more clues of the investment activities than PCC.


\subsection{Discovering the Evolving Co-investment Patterns}

In this experiment, we show the capability of DeepCNL (DNL) to capture the evolving deep co-investment patterns. We learn the annual deep co-investment networks by DNL from the year 2010 to 2013 with $\gamma=0.001$ and extract the biggest connected components from the corresponding networks. The result of DNL is shown in Figure 5 (c-f) and we also show the learned networks (from the year 2010 to 2011) by PCC with the same parameters in Figure 5 (a) and (b) as the comparison.
\begin{figure}[ht]
  \caption{The biggest connected components in the annual deep co-investment networks learned by DNL and PCC with $\gamma=0.001$}
  \centering
    \subfigure[2010 (PCC)]{
    \label{fig:subfig:b} 
    \includegraphics[width=1.6in]{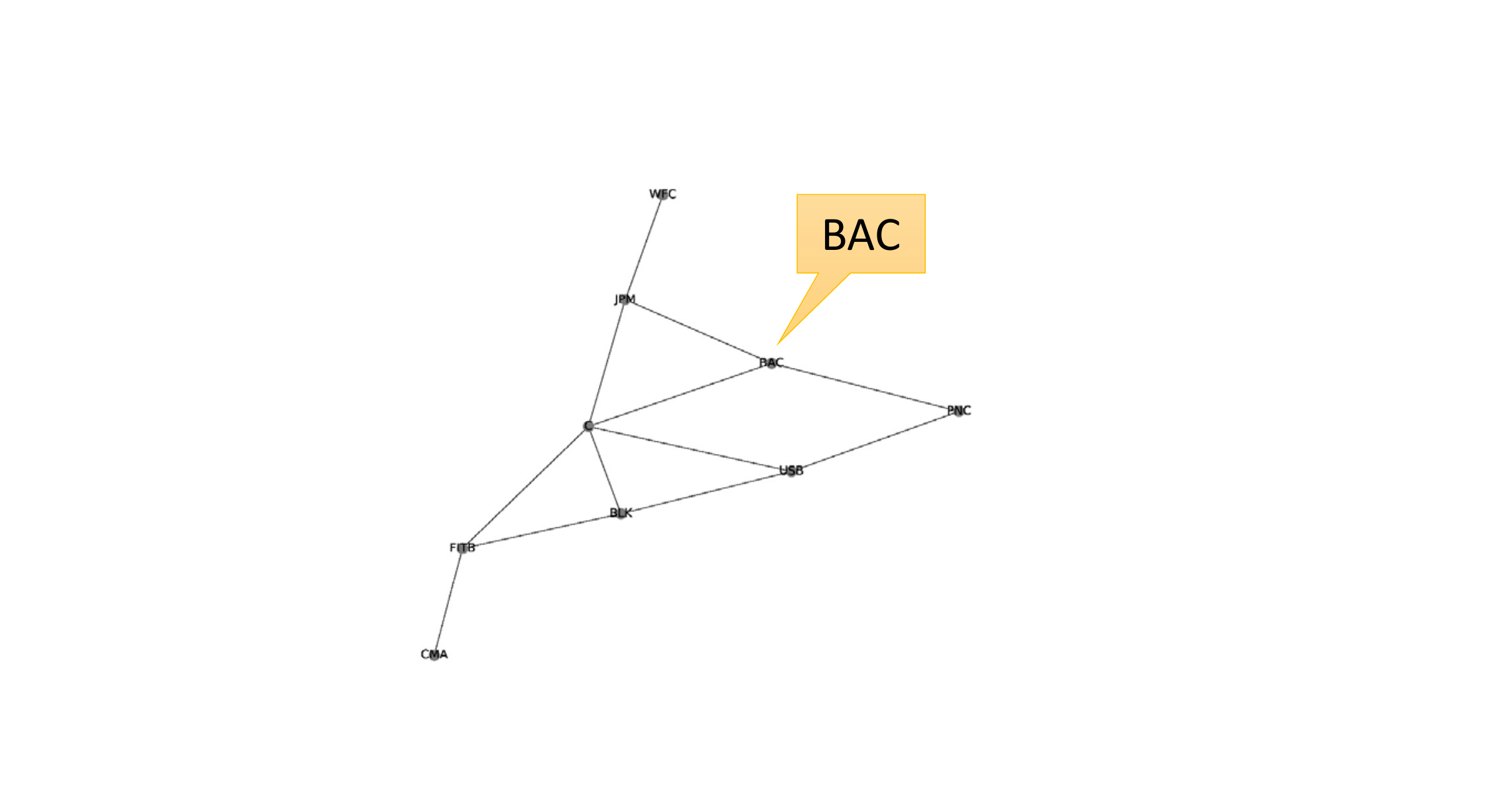}}
      \subfigure[2011 (PCC)]{
    \label{fig:subfig:b} 
    \includegraphics[width=1.6in]{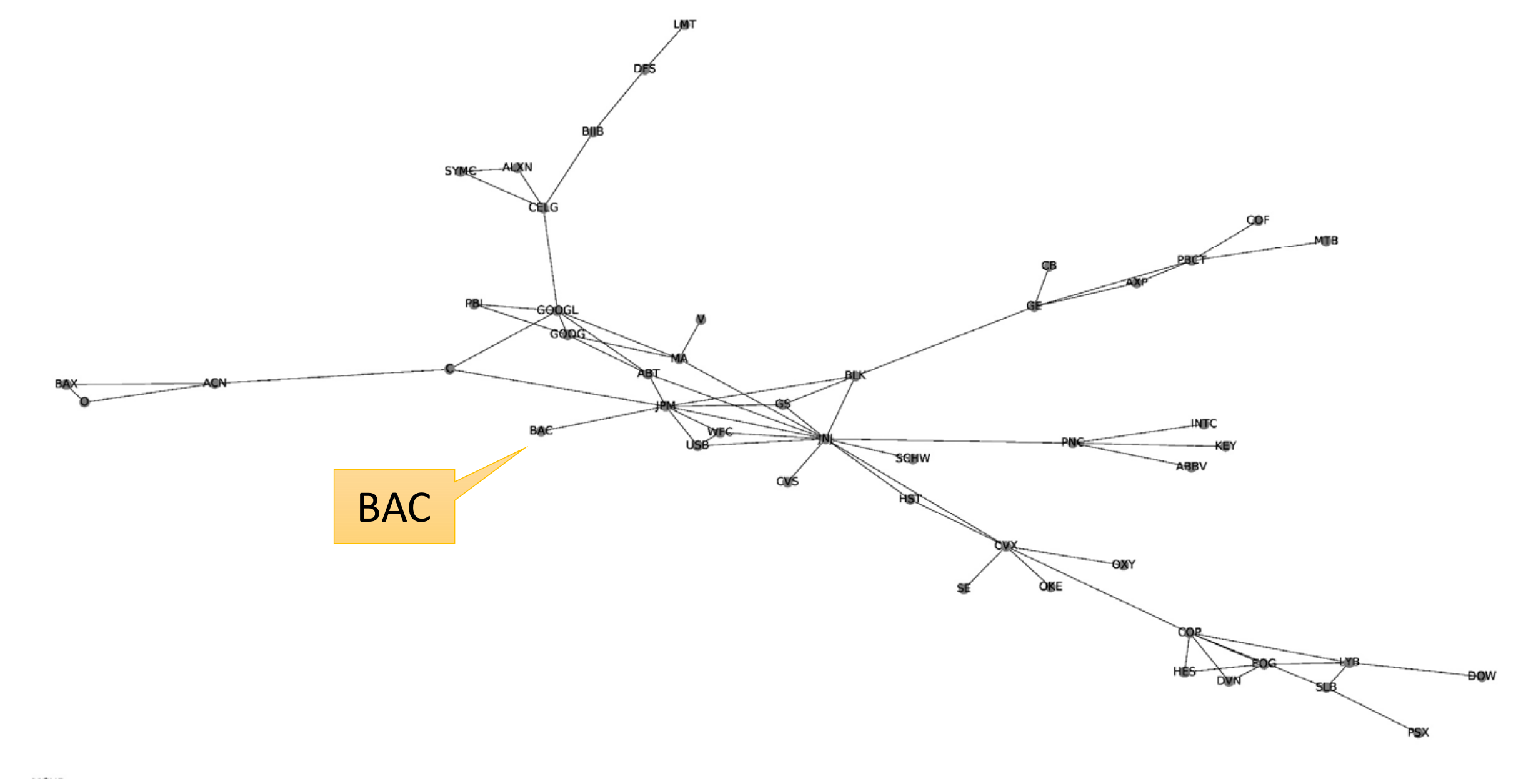}}
  \subfigure[2010 (DNL)]{
    \label{fig:subfig:a} 
    \includegraphics[width=1.6in]{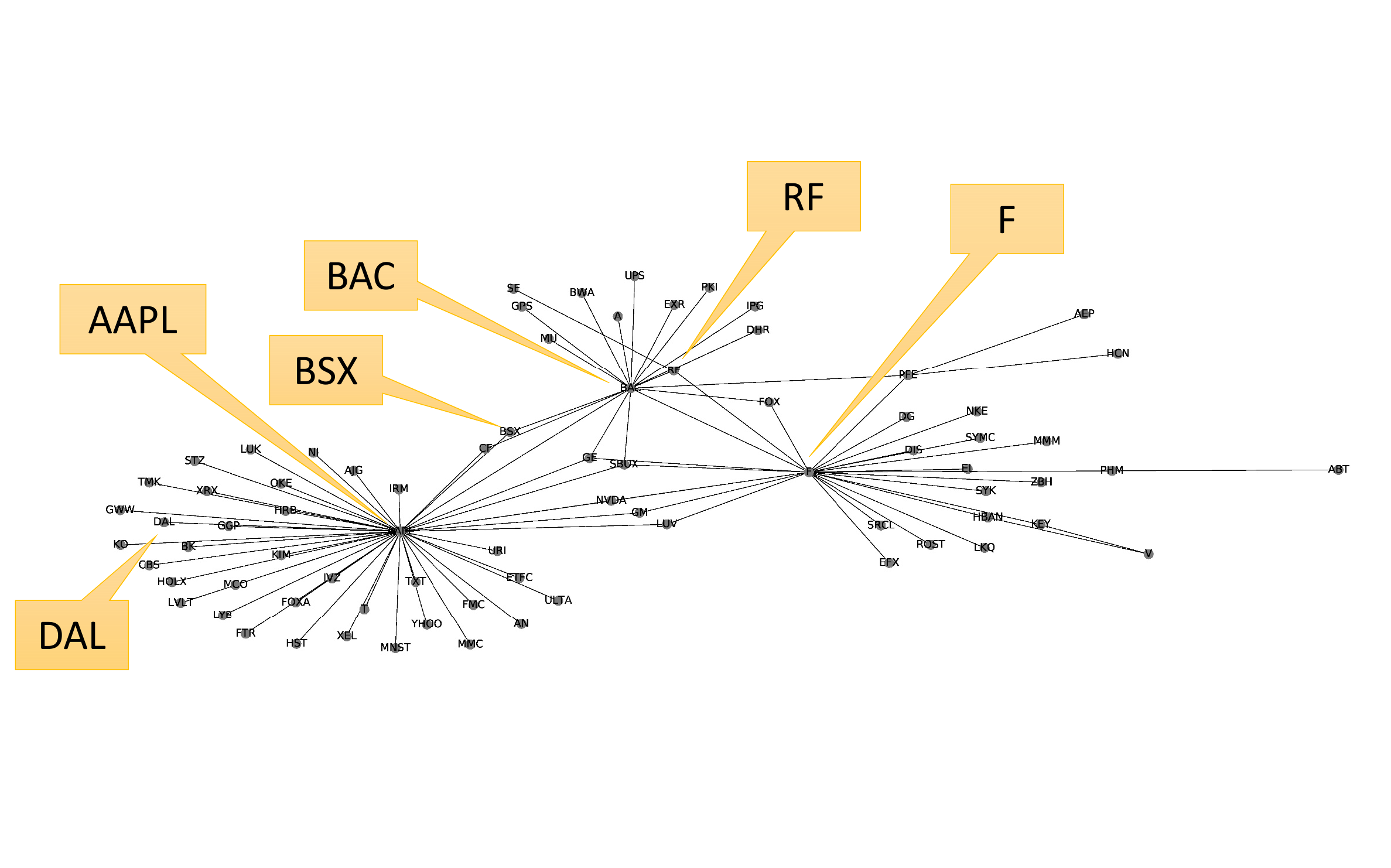}}
  \subfigure[2011 (DNL)]{
    \label{fig:subfig:b} 
    \includegraphics[width=1.6in]{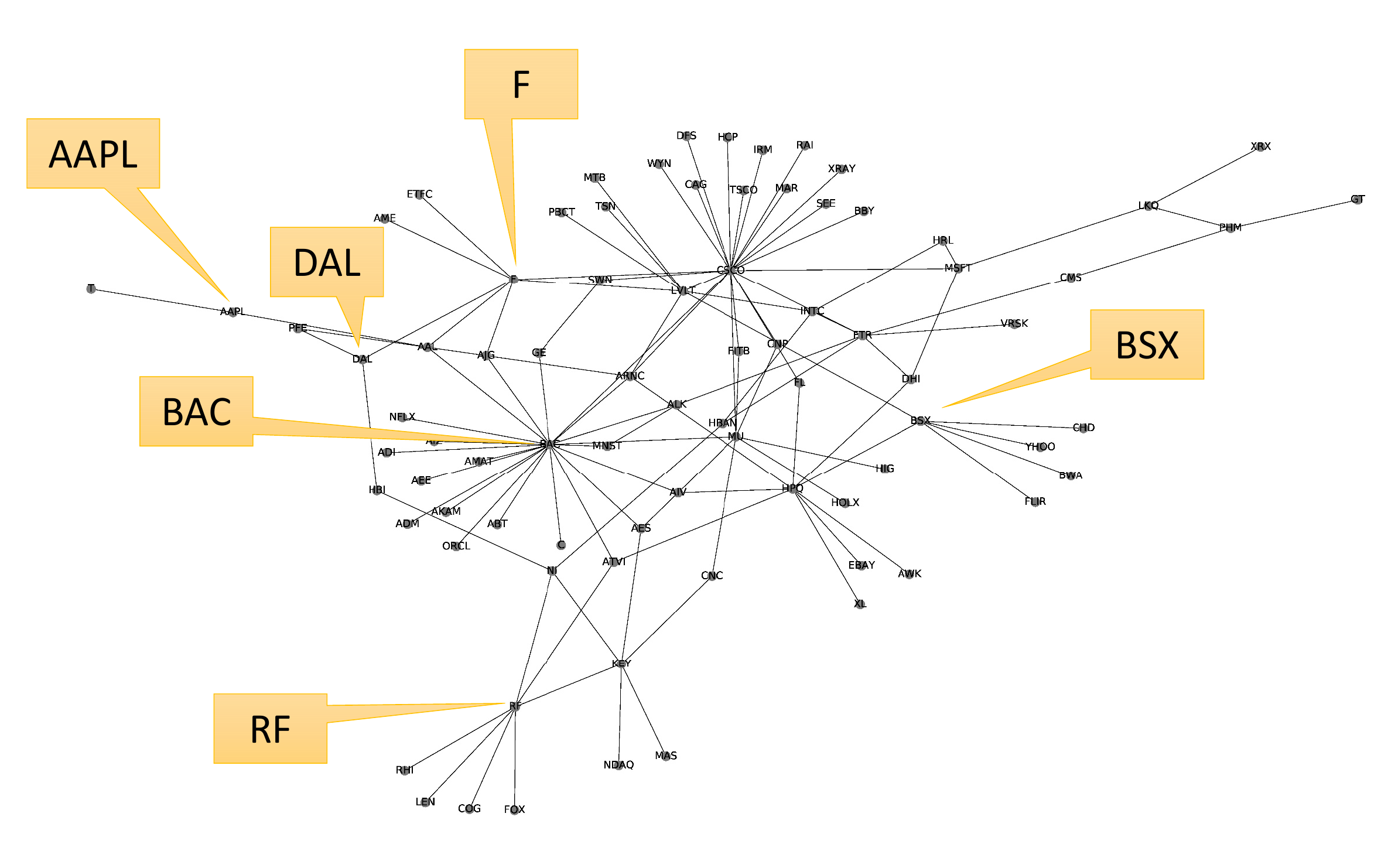}}
     \subfigure[2012 (DNL)]{
    \label{fig:subfig:c} 
    \includegraphics[width=1.6in]{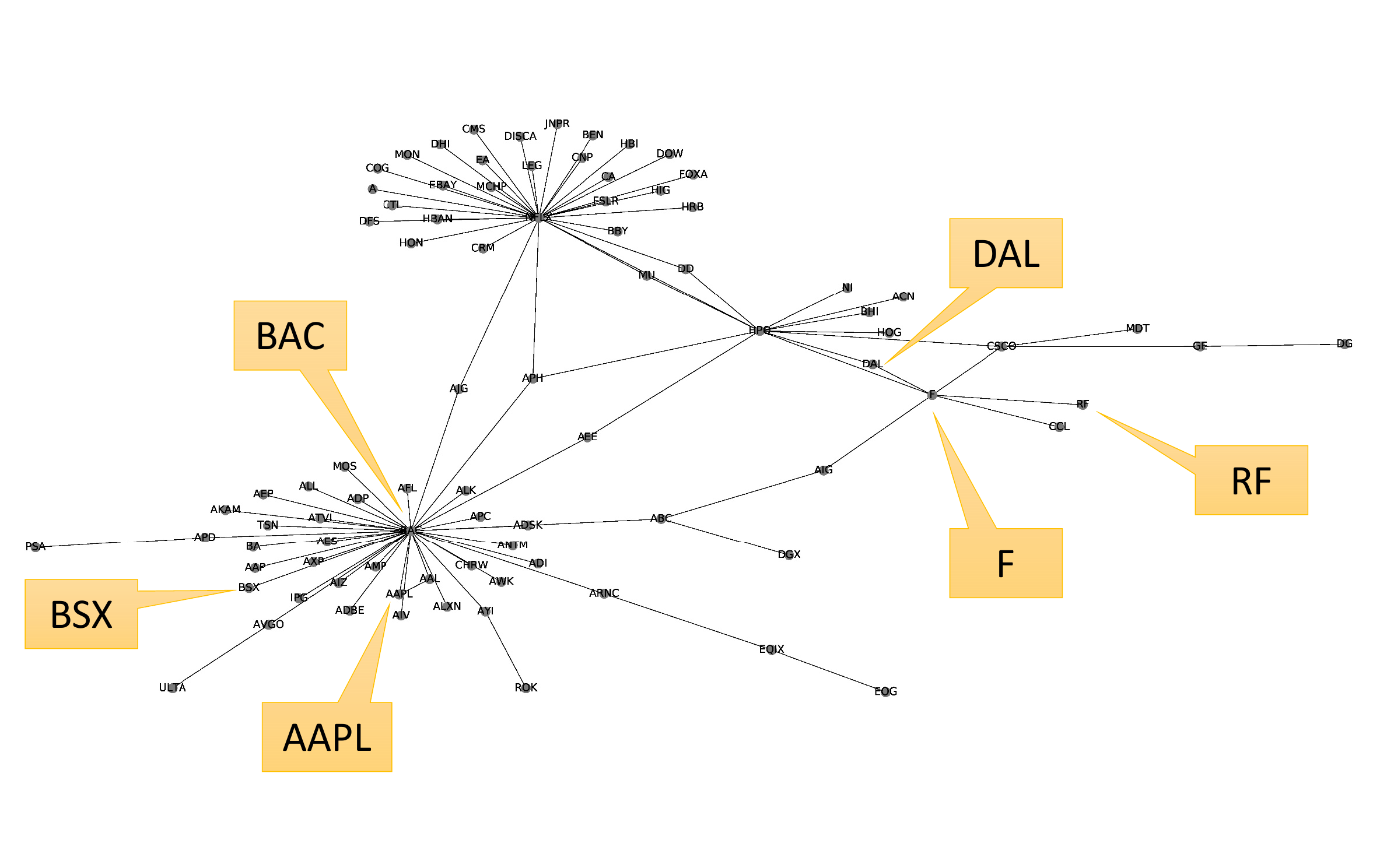}}
     \subfigure[2013 (DNL)]{
    \label{fig:subfig:d} 
    \includegraphics[width=1.6in]{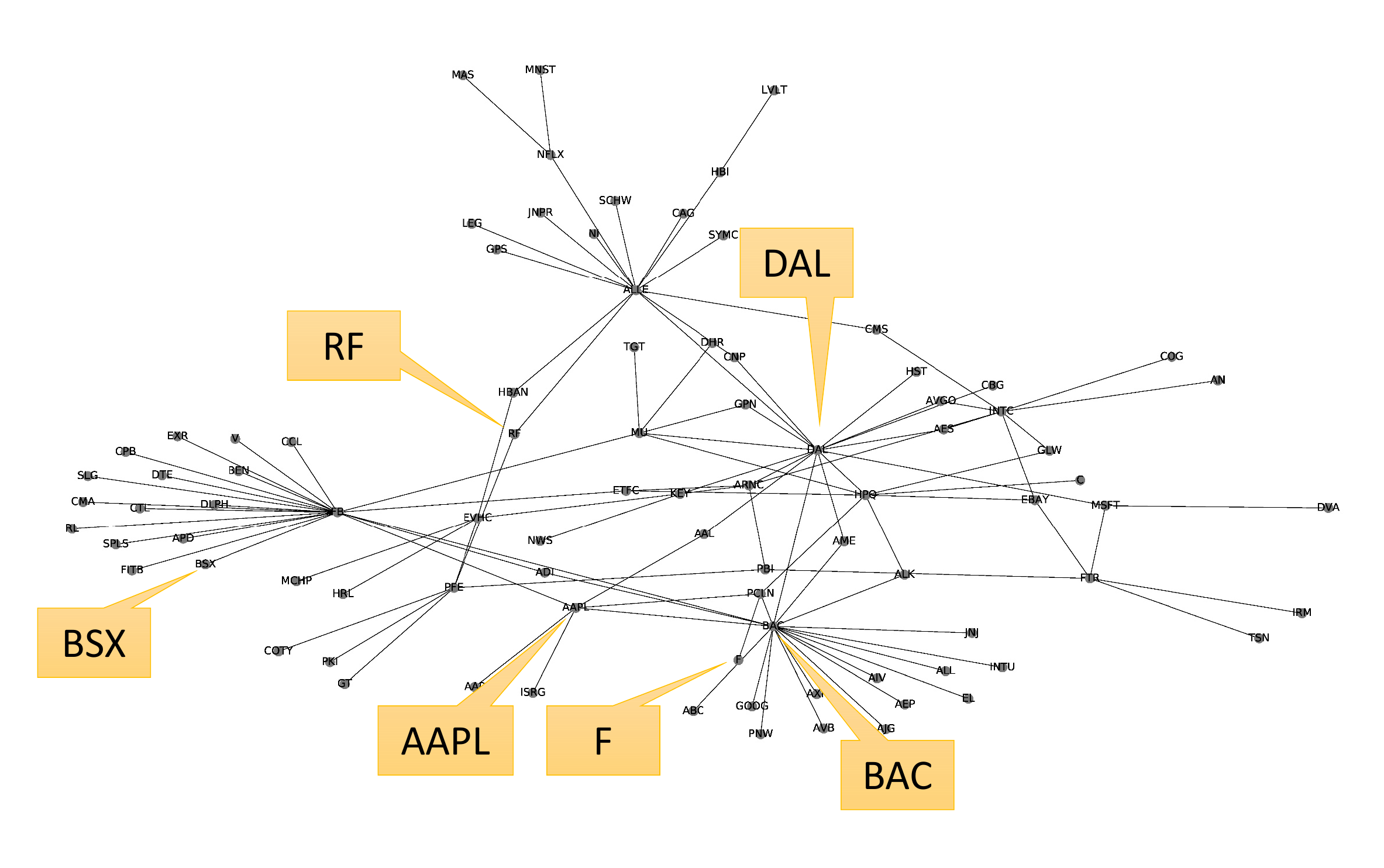}}
  \label{fig:subfig} \vspace{-0.1in}
\end{figure}
We can observe that the obtained connected components by DNL \textbf{keep a relatively stable structure} although the co-investment patterns are evolving annually and this indicates that our method captures the effective evolving co-investment patterns. What's more, we further find out that the \textbf{results of DNL in Figure 5 cover 60\% of the top stocks} mentioned by the news reports found in the related study \cite{10.1371/journal.pone.0146576}. This shows that our result follows the financial rule about the positive correlation between the transaction volume of a stock and the number of times that it is mentioned in the news media \cite{wrap58595}. As a comparison, the results of PCC hardly capture any useful stable structure related to the mentioned studies.

We also observe that the distances between the high degree nodes (stocks) of DNL's results in Figure 5 match the discoveries of several financial studies \cite{Rafael2009} \cite{Gregory2018} \cite{RePEc:bap:journl:170303}. To further analyze the captured co-investment rules by DNL, we compute the yearly average distance between the common high degree nodes (stocks) for the obtained connected components of DNL and list the results in Table $\uppercase\expandafter{\romannumeral3}$. Especially, in Table $\uppercase\expandafter{\romannumeral3}$, we find that the average distance between the ``AAPL'' (Apple) and ``BAC'' (Bank of America) is the shortest with the smallest standard deviation. \textbf{This result coincides the most invested stocks in the mutual funds \cite{Rafael2009}, the most co-occurrences stocks in the financial reviews \cite{Gregory2018} }and the ranking result of the stock correlation study \cite{RePEc:bap:journl:170303}.

\begin{table}[h]\scriptsize
\centering
\caption{The average distances for the common high degree nodes}\vspace{-0.1in}
\scalebox{0.85}[0.85]{
\begin{tabular}{ccccccc}
\toprule
&BAC&AAPL&BSX&DAL&RF&F\\
BAC&&\textbf{1.5$\pm$0.5}&1.8$\pm$0.8&2.0$\pm$1.0&2.5$\pm$1.1&2.0$\pm$0.7\\
AAPL&&&2.5$\pm$1.5&2.5$\pm$1.2&3.8$\pm$0.8&2.5$\pm$0.9\\
BSX&&&&3.3$\pm$0.7&3.8$\pm$1.2&3.3$\pm$0.7\\
DAL&&&&&3.0$\pm$0.7&2.3$\pm$1.3\\
RF&&&&&&2.8$\pm$1.8\\
F&&&&&&\\
\bottomrule
\end{tabular}
}
\end{table}\vspace{-0.1in}

\section{Conclusion}

This work learns the co-investment relations for the stocks in the market with the deep co-investment network learning model. Its main contribution is to learn the co-investment relationships which address the impact of the market index trends and link the practical financial rules to the inner parameters of the proposed deep learning model (DeepCNL). The experimental results on the real-world stock data show that DeepCNL learns the effective deep co-investment network that performs consistently with known financial principles on various tasks. This verifies that our model captures the dynamic relationship between the stock-level index and the market-level index.

\section{Acknowledgements}
This work is supported by the National Natural Science Foundation of China (Grant No.61503422,61602535), the Open Project Program of the National Laboratory of Pattern Recognition (NLPR), the Program for Innovation Research in Central University of Finance and Economics, and Beijing Social Science Foundation (Grant No. 15JGC150).
\bibliographystyle{IEEEtran}
\bibliography{ref}

\begin{thebibliography}{10}
\providecommand{\url}[1]{#1}
\csname url@samestyle\endcsname
\providecommand{\newblock}{\relax}
\providecommand{\bibinfo}[2]{#2}
\providecommand{\BIBentrySTDinterwordspacing}{\spaceskip=0pt\relax}
\providecommand{\BIBentryALTinterwordstretchfactor}{4}
\providecommand{\BIBentryALTinterwordspacing}{\spaceskip=\fontdimen2\font plus
\BIBentryALTinterwordstretchfactor\fontdimen3\font minus
  \fontdimen4\font\relax}
\providecommand{\BIBforeignlanguage}[2]{{%
\expandafter\ifx\csname l@#1\endcsname\relax
\typeout{** WARNING: IEEEtran.bst: No hyphenation pattern has been}%
\typeout{** loaded for the language `#1'. Using the pattern for}%
\typeout{** the default language instead.}%
\else
\language=\csname l@#1\endcsname
\fi
#2}}
\providecommand{\BIBdecl}{\relax}
\BIBdecl

\bibitem{JOFI:JOFI339}
N.~Cetorelli and M.~Gambera, ``Banking market structure, financial dependence
  and growth: International evidence from industry data,'' \emph{The Journal of
  Finance}, vol.~56, no.~2, pp. 617--648, 2001.

\bibitem{DBLP:conf/kdd/JohnsonB15}
N.~Johnson and A.~Banerjee, ``Structured hedging for resource allocations with
  leverage,'' in \emph{Proceedings of the 21th {ACM} {SIGKDD}, Sydney, NSW,
  Australia, August 10-13, 2015}, 2015, pp. 477--486.

\bibitem{DBLP:conf/icdm/SilvaBK16}
D.~F. Silva, G.~E. A. P.~A. Batista, and E.~J. Keogh, ``Prefix and suffix
  invariant dynamic time warping,'' in \emph{{IEEE} 16th International
  Conference on Data Mining, {ICDM} 2016, December 12-15, 2016, Barcelona,
  Spain}, 2016, pp. 1209--1214.

\bibitem{DBLP:conf/kdd/DauK17}
H.~A. Dau and E.~J. Keogh, ``Matrix profile {V:} {A} generic technique to
  incorporate domain knowledge into motif discovery,'' in \emph{Proceedings of
  the 23rd {ACM} {SIGKDD}, Halifax, NS, Canada, August 13 - 17, 2017}, 2017,
  pp. 125--134.

\bibitem{JOFI:JOFI2626}
B.~et~al., ``Order imbalances and stock price movements on october 19 and 20,
  1987,'' \emph{The Journal of Finance}, vol.~44, no.~4, pp. 827--848, 1989.

\bibitem{DBLP:conf/sdm/LeeKBM17}
J.~B. Lee, X.~Kong, Y.~Bao, and C.~M. Moore, ``Identifying deep contrasting
  networks from time series data: Application to brain network analysis,'' in
  \emph{Proceedings of the 2017 {SIAM}, Houston, Texas, USA, April 27-29,
  2017.}, 2017, pp. 543--551.

\bibitem{WARTHER1995209}
V.~A. Warther, ``Aggregate mutual fund flows and security returns,''
  \emph{Journal of Financial Economics}, vol.~39, no.~2, pp. 209--235, 1995.

\bibitem{deeplearning2015}
Y.~B. Yann~LeCun and G.~Hinton, ``Deep learning,'' \emph{Nature}, vol. 521,
  no.~10, pp. 436--444, May 2015.

\bibitem{6738831}
A.~Giusti, D.~C. Ciresan, J.~Masci, L.~M. Gambardella, and J.~Schmidhuber,
  ``Fast image scanning with deep max-pooling convolutional neural networks,''
  in \emph{{IEEE} {ICIP} 2013, Melbourne, Australia, September 15-18, 2013},
  2013, pp. 4034--4038.

\bibitem{LUKOSEVICIUS2009127}
M.~LukoAeviAius and H.~Jaeger, ``Reservoir computing approaches to recurrent
  neural network training,'' \emph{Computer Science Review}, vol.~3, no.~3, pp.
  127 -- 149, 2009.

\bibitem{iet:/content/conferences/10.1049/cp_19991218}
F.~Gers, ``\BIBforeignlanguage{English}{Learning to forget: continual
  prediction with lstm},'' \emph{\BIBforeignlanguage{English}{IET Conference
  Proceedings}}, pp. 850--855(5), January 1999.

\bibitem{DBLP:journals/corr/ChungGCB14}
J.~Chung, {\c{C}}.~G{\"{u}}l{\c{c}}ehre, K.~Cho, and Y.~Bengio, ``Empirical
  evaluation of gated recurrent neural networks on sequence modeling,''
  \emph{CoRR}, vol. abs/1412.3555, 2014.

\bibitem{Bridle1990}
J.~S. Bridle, \emph{Probabilistic Interpretation of Feedforward Classification
  Network Outputs, with Relationships to Statistical Pattern
  Recognition}.\hskip 1em plus 0.5em minus 0.4em\relax Berlin, Heidelberg:
  Springer Berlin Heidelberg, 1990, pp. 227--236.

\bibitem{DBLP:journals/corr/KingmaB14}
D.~P. Kingma and J.~Ba, ``Adam: {A} method for stochastic optimization,''
  \emph{CoRR}, vol. abs/1412.6980, 2014.

\bibitem{NIPS2014_5346}
I.~Sutskever, O.~Vinyals, and Q.~V. Le, ``Sequence to sequence learning with
  neural networks,'' in \emph{Advances in Neural Information Processing Systems
  27}, Z.~Ghahramani, M.~Welling, C.~Cortes, N.~D. Lawrence, and K.~Q.
  Weinberger, Eds.\hskip 1em plus 0.5em minus 0.4em\relax Curran Associates,
  Inc., 2014, pp. 3104--3112.

\bibitem{DBLP:conf/icml/JebaraWC09}
T.~Jebara, J.~Wang, and S.~Chang, ``Graph construction and \emph{b}-matching
  for semi-supervised learning,'' in \emph{Proceedings of the 26th Annual
  International Conference on Machine Learning, {ICML} 2009, Montreal, Quebec,
  Canada, June 14-18, 2009}, 2009, pp. 441--448.

\bibitem{doi:10.3102/10769986027001077}
D.~Thissen, L.~Steinberg, and D.~Kuang, ``Quick and easy implementation of the
  benjamini-hochberg procedure for controlling the false positive rate in
  multiple comparisons,'' \emph{Journal of Educational and Behavioral
  Statistics}, vol.~27, no.~1, pp. 77--83, 2002.

\bibitem{DBLP:conf/kdd/BerndtC94}
D.~J. Berndt and J.~Clifford, ``Using dynamic time warping to find patterns in
  time series,'' in \emph{Knowledge Discovery in Databases: Papers from the
  1994 {AAAI} Workshop, Seattle, Washington, July 1994. Technical Report
  {WS-94-03}}, 1994, pp. 359--370.

\bibitem{Lacasa01042008}
L.~L. et~al., ``From time series to complex networks: The visibility graph,''
  \emph{Proceedings of the National Academy of Sciences}, vol. 105, no.~13, pp.
  4972--4975, 2008.

\bibitem{DBLP:conf/nips/ShervashidzeB09}
N.~Shervashidze and K.~M. Borgwardt, ``Fast subtree kernels on graphs,'' in
  \emph{Proceedings of NIPS held 7-10 December 2009, Vancouver, British
  Columbia, Canada.}, 2009, pp. 1660--1668.

\bibitem{Kempe:2003:MSI:956750.956769}
D.~Kempe, J.~Kleinberg, and E.~Tardos, ``Maximizing the spread of influence
  through a social network,'' in \emph{Proceedings of the Ninth ACM
  SIGKDD}.\hskip 1em plus 0.5em minus 0.4em\relax New York, NY, USA: ACM, 2003,
  pp. 137--146.

\bibitem{10.1371/journal.pone.0146576}
G.~Ranco, I.~Bordino, G.~Bormetti, G.~Caldarelli, F.~Lillo, and M.~Treccani,
  ``Coupling news sentiment with web browsing data improves prediction of
  intra-day price dynamics,'' \emph{PLOS ONE}, vol.~11, no.~1, pp. 1--14, 01
  2016.

\bibitem{wrap58595}
M.~Alanyali, H.~S. Moat, and T.~Preis, ``Quantifying the relationship between
  financial news and the stock market,'' \emph{Scientific Reports}, vol. Volume
  3, p. Article number 3578, December 2013.

\bibitem{Rafael2009}
R.~Solis, ``Visualizing stock-mutual fund relationships through social network
  analysis,'' \emph{Global Journal of Finance and Banking Issues}, vol.~3,
  no.~3, 2009.

\bibitem{Gregory2018}
G.~Kramida, ``Analysis of stock symbol co-occurrences in financial articles,''
  \url{https://wiki.cs.umd.edu/cmsc734_f13/images/9/9f/Analysis_of_Stock_Symbol_Co-occurences_in_Financial_Articles.pdf},
  accessed May 26, 2018.

\bibitem{RePEc:bap:journl:170303}
C.~Shekhar and M.~Trede, ``{Portfolio Optimization Using Multivariate t-Copulas
  with Conditionally Skewed Margins},'' \emph{Review of Economics \& Finance},
  vol.~9, pp. 29--41, August 2017.

\end{thebibliography}

\end{document}